\documentclass[aps,prl,english,preprintnumbers,nofootinbib,floatfix,twocolumn,10pt]{revtex4-1}

\usepackage{amsfonts,amsmath,amssymb}
\usepackage{dcolumn}
\usepackage{graphicx,epsfig}
\usepackage{float}
\usepackage{subfig}
\usepackage{subfloat}
\usepackage[utf8]{inputenc}
\usepackage{hyperref}
\usepackage{babel}

 \newcommand{\be}{\begin{equation}}
 \newcommand{\ee}{\end{equation}}
 \newcommand{\bea}{\begin{eqnarray}}
 \newcommand{\eea}{\end{eqnarray}}

\newcommand{\beq}{\begin{equation}}
\newcommand{\eeq}{\end{equation}}

\def\s{\sigma}

\renewcommand*{\thefootnote}{\fnsymbol{footnote}}

\begin{document}

\title{Sources in the Weyl double copy}
\author{Damien A. Easson,$^{\dag}$, Tucker Manton,$^{\dag}$ and Andrew Svesko$^{\ddag}$}
\affiliation{$^\dag$Department of Physics,  
Arizona State University, Tempe, AZ 85287-1504, USA\\
$^\ddag$Department of Physics and Astronomy, University College London, London WC1E 6BT, UK }

\begin{abstract}\vspace{-2mm}
The Weyl double copy relates exact solutions in general relativity to exact solutions in gauge theory, formulated in the spinorial language. To date, the Weyl double copy is understood and employed only for vacuum spacetimes, and hence only to vacuum gauge theories. In this letter, we propose an extension to the Weyl double copy that provides a systematic procedure for treating gravitational sources. We show this extended Weyl double copy gives a new perspective on the Kerr-Newman black hole and the general class of Petrov type D electro-vac spacetimes.

\end{abstract}

\renewcommand*{\thefootnote}{\arabic{footnote}}
\setcounter{footnote}{0}

\maketitle

\noindent \textbf{Introduction.} Historically, gravity (GR) and Yang-Mills gauge theories are described as formal analogs of each other. The perturbative double copy \cite{Bern:2008qj,Bern:2010yg,Bern:2010ue}, initially inspired by the Kawai-Lewellen-Tye relations \cite{Kawai:1985xq}, provides a mechanism to build multi-loop, higher point scattering amplitudes in (super)gravity from the product of Yang-Mills amplitudes, revealing a fundamental connection between the two theories. Not only has the double copy led to powerful techniques in computing gravity amplitudes at tree-level and beyond (see, \emph{e.g.}, \cite{Bern:2010ue,Bern:2010tq,Carrasco:2011mn,Oxburgh:2012zr,Bern:2013yya}), it offers new insight into the nature of perturbative quantum gravity. The success of  the double copy has also  motivated investigations into a \emph{classical} double copy, where spacetimes are understood as copies of classical Yang-Mills solutions \cite{Neill:2013wsa,Monteiro:2014cda,Luna:2016hge,Luna:2018dpt}. 

There are two versions of an exact classical double copy. The first, the Kerr-Schild double copy, interprets spacetimes with a Kerr-Schild decomposition in terms of classical Yang-Mills solutions \cite{Monteiro:2014cda}. More precisely, let $g_{\mu\nu}$ be a $d$-dimensional spacetime admitting a Kerr-Schild form about Minkowski space $\eta_{\mu\nu}$,  $g_{\mu\nu}=\eta_{\mu\nu}+\phi k_{\mu}k_{\nu}$, with $\phi$ a scalar field over the full spacetime and $k^{\mu}$ a null geodesic with respect to $\eta_{\mu\nu}$ and $g_{\mu\nu}$. The Ricci tensor $R^{\mu}_{\;\nu}$ linearizes with respect to $g_{\mu\nu}$, such that for $g_{\mu\nu}$ stationary, $\partial_{0}g_{\mu\nu}=0$ for timelike coordinate $x^0$, and fixing $k_{0}=1$, the linearized Einstein equations are equivalent to Yang-Mills theory linearized with respect to the gauge field $A_{\mu}\equiv\phi k_{\mu}$, 
\begin{equation}\label{generalRicci}
R^{\mu}_{ \ 0}\propto \partial_{\nu}F^{\mu\nu}=J^{\mu},
\end{equation}
where $F_{\mu\nu}=\partial_{[\mu}A_{\nu]}$ is a Maxwell gauge field strength. The stress tensor $T^{\mu\nu}$ sourcing GR is related to sources $J^{\mu}$ in Maxwell's equations by $J^{\mu}=2(T^{\mu}_{\ 0}-T\delta^{\mu}_{\;0}/(d-2))$. The scalar field also must satisfy a wave equation, \emph{e.g.}, $\frac{1}{2}\nabla^2\phi=R^0_{ \ 0}$ in $d=4$. Here $A_{\mu}$ is the single copy and $\phi$ is the zeroth copy. The Kerr-Schild double copy may be modified to describe a variety of spacetimes, including maximally symmetric black holes and gravitational waves \emph{e.g.}, \cite{Luna:2015paa,Bahjat-Abbas:2017htu,Carrillo-Gonzalez:2017iyj, CarrilloGonzalez:2019gof,Easson:2020esh}. 

A second version of the classical double copy is the Weyl double copy \cite{Luna:2018dpt,Godazgar:2021iae}.  It states that given a completely symmetric Weyl spinor $\Psi_{ABCD}$ satisfying the four-dimensional vacuum Einstein equations for Petrov type D or type N spacetimes, we may construct a single copy Maxwell field strength (spinor) $f_{AB}$ and a zeroth copy $S$ as
\begin{equation}\label{OldWeylDC}
    \Psi_{ABCD}=\frac{1}{S}f_{(AB}f_{CD)}.
\end{equation}
The generally complex scalar field $S$ obeys $\Box^{(0)}S=0$, and $ \nabla^{(0)}_{\mu}F^{\nu\mu}=\nabla^{(0)}_\mu\tilde{F}^{\nu\mu}=0$, where $F_{\mu\nu}$ is the tensorial counterpart of the field strength spinor $f_{AB}=\frac{1}{2}F_{\mu\nu}\s^{\mu\nu}_{AB}$,  with $\s^{\mu\nu}_{AB}$ built from Pauli-spin matrices and spacetime vierbeins, and $\tilde{F}_{\mu\nu}=\frac{\sqrt{-g}}{2}\varepsilon_{\mu\nu\alpha\beta}F^{\alpha\beta}$ is the dual field strength. Given $f_{AB}$, it is straightforward to obtain the field strength tensor $F_{\mu\nu}$ (see, for example, \cite{Penrose:1985bww,Luna:2018dpt}, or equation (A.13) of \cite{Keeler:2020rcv}). The $(0)$-superscript indicates that derivatives are to be taken over the flat background metric, obtained from an appropriate limit of the full spacetime metric associated to $\Psi_{ABCD}$. The Weyl double copy is consistent with the Kerr-Schild double copy, with the Kerr-Schild scalar $\phi$ being a linear combination of the real and imaginary parts of $S$, and resolves ambiguities of the latter \cite{Luna:2018dpt}. The Weyl double copy has been used to build double copies for an algebraically special fluid-dual metric \cite{Keeler:2020rcv} and more recently to construct the double copy of the multipole expansion \cite{Chacon:2021hfe}.
 
Significant progress towards a more fundamental understanding of the Weyl double copy has recently been made. In \cite{White:2020sfn,Chacon:2021wbr}, the Weyl double copy is shown to hold for vacuum spacetimes of arbitrary Petrov type at the linearized level. More recently, investigations into the asymptotic behavior of the Weyl double copy \cite{Godazgar:2021iae,Adamo:2021dfg} have provided further insights into the scope of the role algebraic speciality plays in the story. It is also discussed in \cite{Godazgar:2021iae} how an interpretation of the Kerr-Newman spacetime is still lacking in the literature. There thus seems to be a clear calling for new perspectives on how to treat sources in the Weyl double copy, which is the primary goal of this letter.


\noindent \textbf{Weyl double copy with sources.} To account for external sources, we propose the following extension to the Weyl double copy (\ref{OldWeylDC}). Instead of $\Psi_{ABCD}$ being constructed from a single scalar-gauge theory, we promote it to a sum over $m$ products of scalar-gauge theories:
\begin{equation}\label{NewWeylDC}
    \Psi_{ABCD}=\sum_{n=1}^m\frac{1}{S_{(n)}}f^{(n)}_{(AB}f^{(n)}_{CD)},
\end{equation}
where for $n>1,$ each $S_{(n)}$ and $f^{(n)}_{AB}$ satisfy a sourced wave equation and sourced Maxwell equations. Written in terms of the tensor field strength, 
\begin{equation}\label{sources}
    \nabla_\mu^{(0)}F_{(n)}^{\nu\mu}=J_{(n)}^\nu, \ \ \ \ \Box^{(0)}S_{(n)}=\rho_{(n)}^S.
\end{equation}
We thus refer to (\ref{NewWeylDC}) as the \emph{sourced Weyl double copy}. The $n=1$ term corresponds to (\ref{OldWeylDC}), \emph{i.e.}, the vacuum `part' of the metric. Terms associated to $n=2,3,...,m$ meanwhile correspond to each source, that is, all parameters which contribute non-trivially to the Ricci curvature are assigned an individual single copy product labeled by $n$ for each $n>1.$ The toy black hole example in the following section illustrates this process very clearly for an arbitrary number of sources. For Kerr-Newman and the general electro-vac type D solution, there are two terms in (\ref{NewWeylDC}), the second being produced by the presence of the Maxwell source on the gravity side.

An important aspect of  (\ref{OldWeylDC}) is that $S$ and $f_{AB}$ are uniquely determined by $\Psi_{ABCD}$ \cite{Luna:2018dpt}. A concern with (\ref{NewWeylDC}) is the possibility the $n>1$ terms are not uniquely determined. However, when sources are present, we are guided by the non-trivial Ricci tensor as related to the single copy sources in the Kerr-Schild double copy (\ref{generalRicci}). With the knowledge of the $n^{\text{th}}$ source on the gravity side, we can identify the corresponding $S_{(n)}$ and $f^{(n)}_{AB}$ by solving the respective field equations as in (\ref{sources}). The only non-uniqueness lies in any transformations that leave (\ref{sources}) untouched, which are constant shifts in $S_{(n)}$ and the gauge freedom of the field strength tensor $F^{\mu\nu}_{(n)}$. 

One may also suppose that (\ref{NewWeylDC}) need not be a sum over multiple terms. Indeed for Type D spacetimes, the index structure of the left hand side of (\ref{NewWeylDC}) is a single spinor product, which means the index structure of each term on the right hand side is of the same form. Specifically, $\Psi_{ABCD}=6 o_{(A}o_B\iota_C\iota_{D)}\sum_n\Psi_2^{(n)}$, where $\Psi_2^{(n)}$ is the one nonzero Weyl scalar associated to the $n^{\text{th}}$ source and $\{o_A,\iota_B\}$ form a spinor basis \cite{Penrose:1985bww}. As we detail in \cite{FollowUP}, if we attempt to construct a Weyl double copy of a sourced spacetime with a single product of spinor field strengths, the connection to the Ricci tensor prescribed by the Kerr-Schild double copy in (\ref{generalRicci}) is lost. This leads us to conclude we must treat each contribution to the Weyl spinor separately. 

The procedure of copying each term individually in the classical double copy can be anticipated from the perturbative double copy for amplitudes, where each term in the amplitude corresponding to a particular propagator structure is independently copied \cite{Bern:2008qj}. A similar procedure happens for double Kerr-Schild examples, such as the Taub-NUT solution \cite{Luna:2015paa}; two gauge fields are summed, one associated to the black hole mass and the other to the NUT charge, and individually copied with distinct null vectors. This process is very similar to our approach to the sourced Weyl double copy, although at the level of the fields rather than the field strengths.

\noindent \textbf{Toy static black hole.} Consider an asymptotically Minkowski black hole with line element
\begin{equation}\label{genmet}
    ds^2=-a(r)dt^2+\frac{1}{a(r)}dr^2+r^2d\Omega^2,
\end{equation}
with $a(r)=1+\sum_{n=1}^{m}a_{n}r^{-n}$ for constants $a_{n}$.
The flat limit of (\ref{genmet}) is Minkowski space in spherical-polar coordinates. We can imagine this black hole being sourced by Lagrangian density $\frac{1}{\sqrt{-g}}L=\frac{1}{2}R+\sum_{n=1}^m\mathcal{L}_{(n)}$. The $n=1$ term is the standard (vacuum) Schwarzschild black hole, with $a_1=-2M$.
The $n=2$ term is the Reissner-Nordstr\"om (RN) contribution, 
and the $n=3$ term may arise by including a magnetic monopole from a nonlinear electrodynamic theory, \emph{e.g.}, \cite{Ayon-Beato:2000mjt}.  The metric (\ref{genmet}) is Petrov type D \cite{FollowUP} with Weyl spinor
\begin{equation}\label{CABCD}
    \Psi_{ABCD}=\sum_{n=1}^m\frac{(n+1)(n+2)}{2}\frac{a_n}{r^{n+2}}o_{(A}o_B\iota_C\iota_{D)}.
\end{equation}
The spinor basis obeys 
 $o_{A}\iota^{A}=o_A\varepsilon^{AB}\iota_B=1,$ where $\varepsilon_{AB}$ is the Levi-Civita symbol, with $\varepsilon_{01}=1$.

From (\ref{NewWeylDC}), the $n^{\text{th}}$ scalar and field strengths are 
\begin{equation}\label{Sandf}
    S_{(n)}=\frac{s_n}{a_n r^n}, \ \ \ \ \ f_{AB}^{(n)}=\frac{e_n }{r^{n+1}}o_{(A}\iota_{B)},
\end{equation}
introducing constants $s_n$ and $e_n$. Evaluating the wave equation for each scalar field,
\begin{equation}\label{BoxS}
    \Box^{(0)}S_{(n)}=\frac{s_n n(n-1)}{a_n r^{n+2}}\equiv \rho^{S}_{(n)},
\end{equation}
 defines the zeroth copy sources $\rho^S_{(n)}$. The Maxwell equations associated to each $f^{(n)}_{AB}$ read
\begin{equation}\label{Max}
    \nabla^{(0)}_{\nu}F_{(n)}^{\mu\nu}=\frac{ e_n(n-1)}{r^{n+2}}\delta^\mu_{ \ t}\equiv \rho^e_{(n)}\delta^\mu_{ \ t}, \ \ \ \  \nabla^{(0)}_{\mu}\tilde{F}_{(n)}^{\nu\mu}=0,
\end{equation}
defining the single copy charge densities $\rho^e_{(n)}$. To relate  scalar-gauge theory sources to the gravity sources, we compute the mixed Ricci tensor. Specifically,
\begin{equation}\label{RicciVector1}
    R^\mu_{ \ t}=-\frac{1}{2}\sum_{n=1}^m n(n-1)\frac{a_n}{r^{n+2}}\delta^\mu_{ \ t},
\end{equation}
whose one nonzero entry is proportional to the gravitational energy density $\rho^{\text{grav}}$,
\begin{equation}\label{rhograv1}
 \rho^{\text{grav}}=\sum_{n=1}^m \rho_{(n)}^{\text{grav}}=\sum_{n=1}^m\frac{(n-1)a_n}{r^{n+2}}.
\end{equation}
Whether we  relate the scalar-gauge sources to (\ref{RicciVector1}) or (\ref{rhograv1}), we find a simple pattern at each order in $n$, namely,
\begin{equation}
    \rho_{(n)}^{\text{grav}}=\frac{a_n}{e_n}\rho_{(n)}^e=\frac{a_n^2}{ns_n}\rho_{(n)}^S,
\end{equation}
which is apparent from (\ref{BoxS}), (\ref{Max}), and (\ref{rhograv1}).

Thus, extending the Weyl double copy as (\ref{NewWeylDC}), we can order-by-order, or `source-by-source,' construct the products $\frac{1}{S}f^2$ whose sum converges to the associated Weyl spinor. As anticipated, the $n=1$ term corresponds to the vacuum part, while $n>1$ respectively correspond to the $n^{\text{th}}$ source parametrized by $a_n.$


\noindent \textbf{Kerr-Newman black hole.} The Kerr-Newman (KN) solution describes a rotating black hole with rotation parameter $a$ equipped with electric charge $Q$ \cite{Newman:1965my}. Its line element in Boyer-Lindquist coordinates $\{t,r,\theta,\phi\}$ is
\begin{equation}\label{KerrNewmanMet}
\begin{split} 
    ds^2&= -\frac{\Delta }{\rho^2}(dt-a\sin^2\theta d\phi)^2 +\frac{\rho^2}{\Delta} dr^2+\rho^2d\theta^2 \\
&+\frac{\sin^2\theta }{\rho^2}\Big((r^2+a^2)d\phi-adt\Big)^2,
    \end{split}
\end{equation}
where
\begin{equation}
    \rho^2=r^2+a^2\cos^2\theta, \ \ \ \ \ \ \ \Delta = r^2-2mr+a^2+Q^2.
\end{equation}
The flat limit ($M,Q\to0$) of (\ref{KerrNewmanMet}) is Minkowski space in oblate spheroidal coordinates. The Weyl spinor is \cite{Adamo:2014baa}
\begin{equation}
\begin{split} 
    \Psi_{ABCD}&=-\frac{6M}{(r+ia\cos\theta)^3}o_{(A}o_B\iota_C\iota_{D)} \\
    &  +\frac{6Q^2}{(r+ia\cos\theta)^3(r-ia\cos\theta)}o_{(A}o_B\iota_C\iota_{D)}.
    \end{split}
\end{equation}
We see this is already in a form suggested by (\ref{NewWeylDC}): let the first term correspond to the parameter $M$,
\begin{equation}
    \frac{1}{S_{(1)}}f^{(1)}_{(AB}f^{(1)}_{CD)}=-\frac{6M}{(r+ia\cos\theta)^3}o_{(A}o_B\iota_C\iota_{D)}.
\end{equation}
This term coincides with the vacuum Weyl double copy  \cite{Luna:2018dpt}, where the zeroth and single copies are identified to be
\begin{equation}
    S_{(1)}=-\frac{q^2}{6M}\frac{1}{r+ia\cos\theta}, \;\;  f^{(1)}_{AB}=\frac{q \, o_{(A}\iota_{B)}}{(r+ia\cos\theta)^2},
\end{equation}
both of which satisfy vacuum equations. The $f_{AB}^{(1)}$ describes electric and magnetic fields sourced by a  rotating disc about the $z$-axis with net charge $q$ \cite{Monteiro:2014cda,Israel:1970kp}. 

Since $Q\neq0$, we also have the $n=2$ contribution
\begin{equation}
    \frac{1}{S_{(2)}}f^{(2)}_{(AB}f^{(2)}_{CD)}=\frac{6Q^2 o_{(A}o_B\iota_C\iota_{D)}}{(r+ia\cos\theta)^3(r-ia\cos\theta)}.
\end{equation}
Introducing Maxwell charge $\tilde{q}$, we choose
\begin{equation}
\begin{split} 
    S_{(2)}&=\frac{\tilde{q}^2}{6Q^2}\frac{1}{(r+ia\cos\theta)(r-ia\cos\theta)}, \\
    f^{(2)}_{AB} &= \frac{\tilde{q}}{(r+ia\cos\theta)^2(r-ia\cos\theta)}o_{(A}\iota_{B)},
    \end{split}
\end{equation}
such that 
\begin{equation}\label{BoxS1}
    \Box^{(0)}S_{(2)}=-\frac{\tilde{q}^2}{3Q^2}\frac{(r^2+a^2)+a^2\sin\theta^2}{\rho^6},
\end{equation}
\begin{equation}\label{Max1}
\begin{split} 
    \nabla^{(0)}_{\mu}F_{(2)}^{\nu\mu}&=-\frac{\tilde{q}}{2}\Big( \frac{(r^2+a^2)+a^2\sin\theta^2}{\rho^6}\delta^\nu_{ \ t}+\frac{2a}{\rho^6}\delta^\nu_{ \ \phi}\Big) \\
    &\equiv\rho_{(2)}^e\delta^\nu_{ \ t}+J_{(2)}\delta^\nu_{ \ \phi},
    \end{split}
\end{equation}
along with $\nabla^{(0)}_\mu\tilde{F}_{(2)}^{\nu\mu}=0.$ The rotation parameter introduces a current density $J_{(2)}$ and hence a magnetic field, consistent with the single copy of the Kerr solution \cite{Monteiro:2014cda}. We obtain the single copy of the RN solution when $a\rightarrow 0$.

 Results (\ref{BoxS1}) and (\ref{Max1}) are as anticipated. Indeed,
\begin{equation}\label{Rmu0KN}
    R^\mu_{ \ t}=-8Q^2\Big(\frac{(r^2+a^2)+a^2\sin^2\theta }{\rho^6}\delta^\mu_{ \ t} +\frac{2a}{\rho^6}\delta^\mu_{ \ \phi}\Big)\propto J^\mu_{(2)}, 
\end{equation}
in agreement with our expectations from the sourced Kerr-Schild double copy (\ref{generalRicci}). Regarding the gravitational energy density, recall that the energy momentum tensor associated with a Maxwell field is traceless, thus $R^\mu_{ \ \nu}\propto T^\mu_{ \ \nu}$ and $R^t_{ \ t}\propto -\rho^{\text{grav}}.$ Therefore we have again verified that $\rho^{\text{grav}}\propto\rho_{(2)}^S\propto\rho_{(2)}^e,$ as is clear from (\ref{BoxS1}), (\ref{Max1}), and (\ref{Rmu0KN}).

For the KN solution there is a particularly simple connection to the Kerr-Schild double copy. The Kerr-Schild ansatz has $\phi k_\mu = \phi^Kk_\mu + \phi^Nk_\mu$ with $\phi^K=\frac{2Mr}{\rho^2}$, $\phi^N=-\frac{Q^2}{\rho^2}$, and $k_\mu = (1,\rho^2/(r^2+a^2),0,a\sin^2\theta).$ Now express the single copy gauge field as $A_\mu = A_\mu^K + A_\mu^N$. The field strength built from $A^K_\mu$ is divergenceless and corresponds to $f^{(1)}_{AB}$, while $A^N_\mu$ produces a field strength whose divergence is proportional to (\ref{Rmu0KN}), thus corresponding to $f^{(2)}_{AB}$.



\noindent \textbf{General type D metric.} As a final example, we consider the most general Petrov type D solution to the Einstein-Maxwell equations, originally found in \cite{Plebanski:1976gy}. After a complex diffeomorphism, the metric can be written in double Kerr-Schild form,
\begin{eqnarray}\label{ComplexPD}
    &ds^2&=\frac{1}{(1-pq)^2}\Big[2i(du+q^2dv)dp-2(du-p^2dv)dq \nonumber \\
   &-& \frac{\mathcal{Q}(q)}{p^2+q^2}(du-p^2dv)^2+\frac{\mathcal{P}(p)}{p^2+q^2}(du+q^2dv)^2\Big],
\end{eqnarray}
with 
\begin{equation}
\begin{split} 
    \mathcal{Q}(q)&=k+e^2+g^2 - 2mq+\epsilon q^2-2nq^3 -(k+\Lambda/3)q^4 \\
    \mathcal{P}(p)&=k+2np-\epsilon q^2+2mp^3-(e^2+g^2+k+\Lambda/3)p^4.
    \end{split}
\end{equation}
 The parameters $m, \ n, \ e, \ g,$ and $\Lambda$ respectively correspond to mass, NUT charge, electric and magnetic monopoles, and the cosmological constant, while $\epsilon$ and $k$ are related to rotation and acceleration. Limits of (\ref{ComplexPD}) include, for example, Kerr-Newman-Taub-NUT and the charged, accelerating black hole (C-metric), which can be obtained by diffeomorphisms, and coordinate and parameter rescalings \cite{Griffiths:2005qp}.

 Here we set $\Lambda=0$. 
The metric is flat when $m=n=e=g=0$, and $\epsilon$ and $k$ are arbitrary, leading to a potentially ambiguous flat space limit \cite{Luna:2018dpt}.  We return to this point momentarily. 
The Weyl spinor and Ricci tensor are
\begin{equation}\label{PsiPD}
\begin{split}
    &\Psi_{ABCD}= -6(m+in)\Bigg(\frac{1-pq}{q+ip}\Bigg)^3o_{(A}o_B\iota_C\iota_{D)} \\
    & +6(e^2+g^2)\Bigg(\frac{1-pq}{q+ip}\Bigg)^{3}\frac{1+pq}{q-ip}o_{(A}o_B\iota_C\iota_{D)},
\end{split}
\end{equation}
\begin{equation}\label{RiccPD}
    R^\mu_{ \ u}=\frac{(e^2+g^2)(p q-1)^4 }{\left(p^2+q^2\right)^3}\Big[\left(p^2-q^2\right)\delta^\mu_{ \ u} +
     2\delta^\mu_{ \ v}\Big],
\end{equation}
where here the timelike coordinate is taken to be $x^0=u$. The stress tensor again corresponds to a Maxwell field and is traceless, with $R^\mu_{ \ \nu}\propto T^\mu_{ \ \nu}$ and $\rho^{\text{grav}}\propto -R^u_{ \ u}$. The first line in (\ref{PsiPD}) corresponds to the vacuum solution studied in \cite{Luna:2018dpt}, where
\begin{equation}\label{f1S1PD}
\begin{split}
    S_{(1)}&=\frac{i}{6}\frac{(\tilde{m}+i\tilde{n})^2(1-pq)}{(m+in)(p-iq)}, \\
&f^{(1)}_{AB}=\frac{(\tilde{m}+i\tilde{n})(1-pq)^2}{(p-iq)^2}o_{(A}\iota_{B)},
     \end{split}
\end{equation}
introducing free parameters $\tilde{m}$ and $\tilde{n}$.

 We can consider (\ref{NewWeylDC}) as having three terms, the second and third coming from $e$ and $g$, though due to the identical coordinate dependence multiplying both parameters in (\ref{PsiPD}), we take the second line of (\ref{PsiPD}) as a single contribution. Thus,
\begin{equation}
\begin{split}
    S_{(2)}&=\frac{(\tilde{e}^2+\tilde{g}^2)^2}{6(e^2+g^2)} \Bigg(\frac{1-pq}{q+ip}\Bigg)\frac{1+pq}{q-ip}, \\
    f^{(2)}_{AB}&=(\tilde{e}^2+\tilde{g}^2)\Bigg( \frac{1-pq}{q+ip}\Bigg)^2\frac{1+pq}{q-ip}o_{(A}\iota_{B)},
    \end{split}
\end{equation}
introducing gauge parameters $\tilde{e}$ and $\tilde{g}$, which we assume are real. Evaluating $\Box^{(0)}S_{(2)}$ leads to
\begin{equation}
\begin{split} 
    \Box^{(0)}S_{(2)}&=\frac{(\tilde{e}^2+\tilde{g}^2)^2}{3(e^2+g^2)}\Bigg(\epsilon\frac{(1-pq)^4(p^2-q^2)}{(p^2+q^2)^3} \\
    &+2k\frac{(1-pq)^4(1+p^2q^2)}{(p^2+q^2)^3} \Bigg)\\
    &\propto \epsilon\rho^{\text{grav}}+k\Delta (p,q),
    \end{split}
\end{equation}
where $\Delta(p,q)$ is the function multiplying the constant $k$. This shows our expectation that $\rho^S_{(2)}\propto\rho^{\text{grav}}$ is met only if we define the flat background with $k=0,$ which is not necessarily problematic; the vacuum equations for $S_{(1)}$ and $f^{(1)}_{AB}$ are immune to the values of $\epsilon$ and $k$. Moreover, if $\epsilon=0,$ then $\Box^{(0)}S_{(2)}=0,$ which is inconsistent with the Kerr-Schild double copy. This suggests the sourced Weyl double copy can specify an appropriate flat background if the choice is ambiguous. 

Finally, for the field strength, it follows
\begin{equation}\label{uvMax}
\begin{split}    
\nabla^{(0)}_\nu F_{(2)}^{\mu\nu}&=-\frac{(\tilde{e}^2+\tilde{g}^2)(p q-1)^4}{2\left(p^2+q^2\right)^3}\left[(q^2-p^2)\delta^\mu_{ \ u} + 2\delta^\mu_{ \ v}\right].
    \end{split} 
\end{equation}
Comparing to (\ref{RiccPD}), we see $R^\mu_{ \ u}\propto J^\mu_{(2)}$. Meanwhile, the dual field strength has a nontrivial divergence
\begin{equation}\label{DivFdual}
\begin{split}
\nabla^{(0)}_\nu\tilde{F}_{(2)}^{\mu\nu}&=-\frac{(\tilde{e}^2+\tilde{g}^2)(1-pq)^4}{2(p^2+q^2)^3}\left[2p^2q^2 \delta^{\mu}_{ \ u}+(q^2-p^2)\delta^\mu_{ \ v}\right],
\end{split}
\end{equation}
pointing to the presence of a magnetic monopole and current. This is in fact expected. The Taub-NUT spacetime studied in \cite{Luna:2015paa} using the Kerr-Schild double copy showed the NUT parameter was found to map to a magnetic monopole in the single copy gauge theory. In our case there is also non-zero angular momentum, which results in the single copy having a current density. Since here we have both a NUT parameter and angular momentum, both a magnetic monopole and magnetic current should be present in the single copy gauge theory. Note the divergence (\ref{DivFdual}) vanishes when taking the limit from the general solution (\ref{ComplexPD}) to Kerr-Newman (\ref{KerrNewmanMet}) \cite{Plebanski:1976gy}.

\noindent \textbf{Discussion.} We have presented a simple extension to the Weyl double copy \cite{Luna:2018dpt} which accounts for external sources in a fashion consistent with the Kerr-Schild double copy. In each example, energy density on the gravity side maps to charge density on the gauge side. This is in line with the intuition from the single copy of the Schwarzschild black hole, where the point mass is mapped to the point charge. Here, a similar process occurs `source-by-source.' When rotation is present, $T^\mu_{ \ \nu}$ has entries corresponding to angular momentum, which appears in the single copy gauge theory as current density and hence produces a magnetic field. If the gravity solution also has a NUT parameter, a magnetic monopole and magnetic current will be present in the single copy.


A full understanding of how sources should be accounted for in the Weyl double copy should necessitate a prescription for the Ricci spinor as well as for the fields that produce the source, which we have not provided here. We explore these relations in an accompanying work \cite{FollowUP}, where we give an explicit copying procedure for the source in pure Einstein-Maxwell theory involving the fields $S_{(1)}, \ S_{(2)},$ and $f_{AB}^{(2)}$. We also study various limits of the general type D metric including the charged C-metric and the Kerr-Newman-Taub-NUT. 

Incorporating sources into the Weyl double copy paradigm allows for many unanswered questions to be explored. In \cite{Monteiro:2020plf}, the (vacuum) Weyl double copy was shown to arise naturally from the double copy of three-point amplitudes; the amplitudes of pure Einstein-Maxwell theory are not currently known to arise from a double copy.  Motivated by (\ref{NewWeylDC}), one possibility is an analogous linearization scheme for including sources in field theoretic scattering processes. Moreover, combining the double copy with amplitude techniques allows for a sophisticated approach towards obtaining the conservative potential for binary black hole systems (\emph{e.g.} \cite{Bern:2019nnu,Bern:2021dqo,Brandhuber:2021eyq}). With the inclusion of sources, it may be possible to extend current results to account for charged, rotating black holes in scattering or merger processes. Having a consistent single copy for the Kerr-Newman black hole is a step in the right direction.

In future work, we intend to investigate spacetimes whose Weyl spinor does not simply decompose into a sum of vacuum plus source terms, suggesting (\ref{NewWeylDC}) needs to be modified. Perhaps the more fundamental twistor methods \cite{White:2020sfn,Chacon:2021wbr,Adamo:2021dfg} will point to a deeper explanation for (\ref{NewWeylDC}), and its potential modification. Lastly, it would be interesting to generalize (\ref{NewWeylDC}) so that it may be applied to sourced spacetimes of different Petrov types. Such a result would set the stage for a double copy procedure for nearly all exact solutions in GR.

\noindent \emph{Acknowledgements.} We are grateful to Andr\'es Luna, Ricardo Monteiro, and Chris White for useful correspondence. DE is supported by a grant from FQXi. AS is supported by the Simons Foundation via the \emph{It from Qubit collaboration}.

\bibliographystyle{apsrev4-1}
\bibliography{references1}

\end{document}